\documentclass[
aps,%
12pt,%
final,%
notitlepage,%
oneside,%
onecolumn,%
nobibnotes,%
nofootinbib,%
superscriptaddress,%
noshowpacs,%
centertags]%
{revtex4}

\begin{document}

\selectlanguage{english}

\author{A.K. Likhoded}
\email{Likhoded@ihep.ru}
\affiliation{Institute for High Energy Physics, Protvino, Russia}
\author{A.V. Luchinsky}
\email{Alexey.Luchinsky@ihep.ru}
\affiliation{Institute for High Energy Physics, Protvino, Russia}
\title{Charmonium production in hadronic experiments at the energy 70 GeV}
\begin{abstract}
The production of the charmonium states in hadronic experiments is considered in NLO at the strong coupling constant. It is shown, that such an approach solves some significant problems, arising when only leading order processes are considered. In particular, in such a consideration  distributions over the transverse momentum of the final charmonium can be obtained. There appears also a natural explanation to the existence of $\chi_{c1}$-meson in final state, that is observed experimentally and cannot be produced in leading order processes.
\end{abstract}

\maketitle

\section{Introduction}

It is well known, that the specific property of the charm production processes is high sensitivity to the gluonic contents of the interacting particles. This property is used not only in the determination of gluon distribution in the proton $G(x)$ \cite{Gallo:2005sh}, but also, in the case of polarized beams, for the determination of the gluon polarization fraction $\Delta G/G$ \cite{Abramov:2005mk}. That is why new experiment, labeled as SPASCHARM, on hadronic production of charmonia was recently proposed. The goal of this experiment is detailed analysis of charmonia production in polarized proton collisions at the 70 GeV energy.

The important part of the Program is the detailed analysis of all possible mechanisms of charmonia production. Such an analysis is especially important since at low energies the contributions of gluon-gluon, quark-gluon and quark-antiquark subprocesses are comparable. For example, if the energy of the proton beam is equal to 40 GeV, the ratio of $\JP$ production cross sections in $p\bar p$ and $pp$ collisions equals $\sigma(p\bar p)/\sigma(pp)\sim6$, so it is clear that the quark-antiquark annihilation plays an important role. As the energy rises this ratio tends to unity and the gluon-gluon process dominates.

Another problem is that the direct production of $\JP$-meson is suppressed in comparison with the production of the intermediate $P$-wave states $\chi_{c0,1,2}$ with the subsequent decay $\chi_c\to J/\psi\gamma$. In the experiments \cite{Alexopoulos:1999wp} this fact is confirmed with one  important remark. The experimentally observed cross sections of $\chi_{c2}$ and $\chi_{c1}$ are comparable ($\chi_{c0}$-meson can hardly be observed due to its small radiative width), while the well known Landau-Yang theorem forbids the formation of the axial meson from two massless gluons. Thus, the experimental observation of this state shows that virtual gluons give additional contribution that is not accounted in the usual description of gluon distributions. There are also large uncertainties in the available parameterizations of these distributions in the region of small virtualities and the values of the gluon momentum fraction $x_g\in[0.1 ; 0.5]$ \cite{Gallo:2005sh}. Due to these uncertainties large errors in the prediction of the $\JP$ and $\chi_c$ production cross sections appear. One more difficulty is that the distributions $G(x)$ are integrated over the transverse momentum. As a result, such method does not allow one to obtain the distributions of $\chi_{c0}$- and $\chi_{c2}$-mesons over $p_T$.

Initially this problems were solved by the introduction of color-octet (CO) components of the quarkonia, that arise naturally in the non-relativistic quantum chromodynamics (NRQCD), where the expansion over the relative velocity of quarks in meson is performed. In this model it is assumed, that final meson is formed from heavy quark pair in color-octet state, that subsequently transforms into a physically observed colorless meson. In the framework of NRQCD the probabilities of these transitions are described by the matrix elements of four-fermion operators, that are determined from the experimental distributions over the transverse momentum of final charmonium. In the works \cite{Cho:1995vh,Braaten:2000cm} it was shown, that octet components give satisfactory agreement with the experimental data on TEVATRON at the energy $\sqrt{s}=1.8$ TeV. We would like to stress, however, that this explanation will not work for charmonium production at lower energies. The reason is that the distributions caused by octet components decreases slowly with the rise of the transverse momentum, but the probability to find such a component in the meson is small, compared with the singlet case. As a result, in the large transverse momentum region the contribution of octet components can be significant, but for small energies and transverse momenta it is suppressed.

Recently another way to solve this problem was proposed, where the so called non-integrated over the transverse momentum distribution functions $G(x,k_T)$ are used ($k_T$-factorization) \cite{Teryaev:1996sr,Kniehl:2006sk,Baranov:1}. In this case both mentioned above problems are solved simultaneously . The transverse momentum of the produced in gluon fusion $\chi_{c0,2}$ mesons is explained by the transverse momenta of the initial partons. Axial charmonium meson can also be produced in gluon fusion, since in the framework of $k_T$ factorizations gluons have non-zero virtuality of the order $k_T^2$. There is a number of works, that explain the experimental distributions on TEVATRON with the help of these functions (see, for example \cite{Hagler:2000dd,Likhoded:2006xu,Kniehl:2006sk}). According to these works, there is no need to introduce CO components to reproduce the experimental data on $P$-wave charmonium $p_T$ distributions. Thus, in $k_T$-factorization approach color-singlet components give the main contribution.

Unfortunately, the method, used in the modeling of the unintegrated distribution functions $G(x,k_T)$, is based on the summation of large $\log(1/x)$, so it is not applicable for low energies, where the gluon momentum fractions are in the range $0.1<x_g<0.5$. For this reason we are forced to use the following approximation in our calculations. We start from the collinear gluon distributions with well known collinear distribution functions. Further we consider the charmonia production at next to leading (NLO) order in the strong coupling constant $\alpha_s$. Such a trick enables us to obtain the distributions over $p_T$ for all charmonium states. For $\chi_{c0}$ and $\chi_{c2}$ production we observe a collinear singularity at $p_T=0$. To avoid this singularity we introduce a cutoff on the transverse momentum, and the value of the cutoff parameter is determined from the inverse geometrical size of the charmonium meson. For directly produced $\JP$ and $\chi_{c1}$ such a singularity is absent and we use the whole integration region for $p_T$.

In the next section we will briefly describe the formalism used in the subsequent paper. Section 3 is devoted to the consideration of different modes of charmonia production and analytical expressions for corresponding partonic cross sections are presented. In the fourth section we determine the cross sections of the hadronic processes and present numerical results. In the last two sections we give our estimates for spin asymmetries and briefly discuss the results of the paper.

\section{Charmonia}

One of the characteristic features of charmonium states is the smallness of the relative quark velocity $v$:
\begin{eqnarray*}
v^2 &\sim& \left[\alpha_s(m_c v)\right]^2\approx 0.2.
\end{eqnarray*}
Due to the existence of this small parameter the processes of charmonia production actually evolves in two almost independent steps:
\begin{itemize}
\item  hard process, in which the quark-antiquark pair is created,
\item subsequent hadronization of this pair into an experimentally observable meson.
\end{itemize}
Second characteristic feature of charmonia is the smallness of the strong coupling constant at the hard step of their production. So it is possible to use the perturbation theory while calculating the hard part of the amplitude.

The second stage of charmonia production, i.e. the hadronization of the quark-antiquark pair into a final meson with the momentum $p$ and mass $M$ \footnote{
In what follows we will neglect the difference between $\JP$ and $\chi_c$ masses
}, is described by the following simple procedure \cite{Braaten:2002fi}. First we use the common QCD to write the amplitude of the production of on-shell quark and antiquark with momenta $p+q/2$ and $p-q/2$ respectively:
\beqa
\M^\mathrm{hard} &=& \bar u\left(\frac{p}{2}+q\right)\M v\left(\frac{p}{2}-q\right).
\eeqa
Here $\bar u$ and $v$ are the wave functions of quark and antiquark, $p$ is the momentum of the final meson ($p^2=M^2$), and $q$ is the relative momentum of quark-antiquark pair ($pq=0$). Then we project this amplitude to the color singlet state with the value of the total spin $S=1$:
\beqa
v_i\left(\frac{p}{2}-q\right) \bar u_j\left(\frac{p}{2}+q\right) &\to& \frac{1}{2\sqrt{6}M^2}\left\{
  \left(\frac{\hat p}{2}-\hat q+m\right)\hat\eps_S(\hat p+m)\left(\frac{\hat p}{2}+\hat q+m\right)
\right\}_{ij},
\eeqa
where $i$ and $j$ are spinor indices of quark and antiquark, $m=M/2$ is $c$-quark mass, and $\eps_S$ is spin polarization vector of the pair. The resulting amplitude is then expanded into a series in relative momentum $q$:
\beq
\M &=& \eps_S^\mu\left( \M_\mu+\M_{\mu\nu}q^\nu+\dots\right).
\label{eq:amp}
\eeq
The amplitude of the $S$-wave charmonium meson production can be derived from the first term of this expansion:
\beqa
\M(\JP) &=& \frac{|R_\psi(0)|}{\sqrt{4\pi m}}\M_\mu \eps^\mu ,
\eeqa
where $\eps_\mu$ is the polarization vector of $\JP$-meson, and $R_{\JP}(0)$ is the radial part of its wave function at the origin. Amplitudes of $P$-wave charmonia production are derived from the second term of the expansion \eqref{eq:amp}:
\beqa
\M(\chi_{c0}) &=& \frac{1}{\sqrt{3}}\frac{|R'_\chi(0)|}{4\pi m} \M_{\mu\nu}J^{\mu\nu}, \\
\M(\chi_{c1}) &=& \frac{i}{2\sqrt{2}m}\frac{|R'_\chi(0)|}{4\pi m} \M_{\mu\nu}\eps_{\rho\sigma\mu\nu}p^\rho\eps_\chi^\sigma,\\
\M(\chi_{c2}) &=& \frac{|R'_\chi(0)|}{4\pi m} \left[
  \frac{J^{\rho\mu}J^{\sigma\nu}+J^{\rho\nu}J^{\sigma\mu}}{2}-\frac{1}{3}J^{\rho\sigma}J^{\mu\nu}
\right] \eps_\chi^{\rho\sigma}\M_{\mu\nu}.
\eeqa
Here $R'_\chi(0)$ is the derivative of the radial part of $\chi$-meson wave function at the origin, $\eps_\chi^\sigma$ and $\eps_\chi^{\rho\sigma}$ are polarization vector and tensor of $\chi_{c1}$- and $\chi_{c2}$-mesons, and the tensor $J^{\mu\nu}$  is
\beqa
J^{\mu\nu} &=& \frac{p^\mu p^\nu}{M^2}-g^{\mu\nu}.
\eeqa

The value of radial part of charmonium wave function can be determined by solving the Schrodinger equation, or from the experimental values of the decay widths of these mesons. In our paper we will use the last way. $\JP$-meson wave function is determined from its leptonic decay width:
\beqa
\Gamma(\JP\to\epem) &=& \frac{4\pi}{3}e_c^2\alpha^2 \frac{|R(0)|^2}{M^2}.
\eeqa
The derivative of the radial part of $\chi_{c0,2}$-meson wave function can be determined form the total widths of these mesons, that are approximately equal to widths of the decays into a pair of massless gluons:
\beq
\Gamma(\chi_0) &=& 96\alpha_s^2 \frac{|R'(0)|^2}{M_\chi^4},\label{eq:G0}\\
\Gamma(\chi_2) &=& \frac{128}{5}\alpha_s^2 \frac{|R'(0)|^2}{M_\chi^4}\label{eq:G2}.
\eeq
The ratio of these width
\beqa
\frac{\Gamma(\chi_{c0})}{\Gamma(\chi_{c2})} &=& \frac{15}{4},
\eeqa
agrees well with the experimental data.

It should be noted, that in the described above procedure we have neglected the relative motion of quarks in the charmonium and use the Bethe-Heitler approximation. Recently it was shown \cite{Liu:2002wq,Liu:2004ga,Bondar:2004sv,Braguta:2005gw,Braguta:2006nf}N that relative motion of quarks results to significant increase of exclusive cross section. It is still not clear, whether it is valid to neglect this motion in the calculation of inclusive cross sections.

The other approximation used in our paper is the CS-approximation \cite{Kartvelishvili:1978id,Gershtein:1980jb,Berger:1980ni,Baier:1981uk}. In other words we have considered only CS contributions and neglected CO components of charmonium mesons. As it was mentioned in the introduction, we think that such an approach can be used, because  the experimental data on the charmonium production on TEVATRON is reproduced in the framework of this model \cite{Kniehl:2006sk,Hagler:2000dd,Hagler:2000eu}.

\subsection{$gg\to\chi_c$}

\begin{figure}
\includegraphics{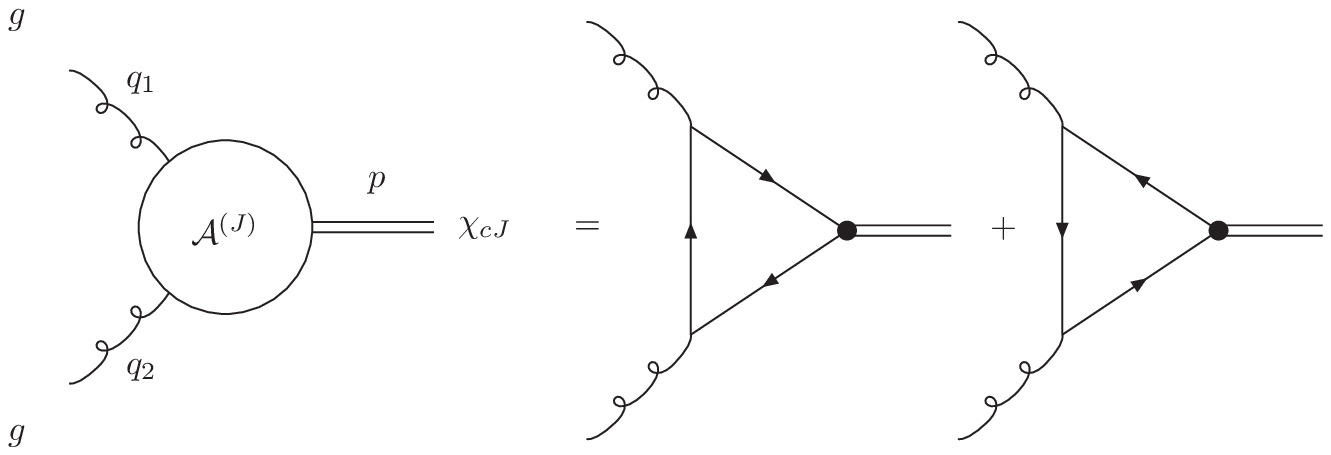}
\caption{$gg\to\chi_{cJ}$\label{diag:Chi}}
\end{figure}

In the following we will need the expressions for effective vertex, that describes the transition of gluon pair into a $\chi_{cJ}$ state. The corresponding diagrams are shown on fig. \ref{diag:Chi}. Using the presented above technique we have derived analytical expressions for these vertices. In the case of scalar meson, for example, we have
\beq
\A^{(0)}_{\mu\nu} &=& \frac{4\sqrt{\pi}\alpha_s|R'_\chi(0)|}{\sqrt{3M^3}(q_1q_2)^2}\left\{
(3M^2-q_1^2-q_2^2)q_2^\alpha q_1^\beta+(M^2-q_1^2-q_2^2)q_1^\alpha q_2^\beta - 2q_2^2 q_1^\alpha q_2^\beta- 2q_1^2 q_2^\alpha q_2^\beta -
\right.\nonumber \\ &-& \left.
\frac{g^{\alpha\beta}}{2}\left(
3M^4-4(q_1^2+q_2^2)M^2+(q_1^2-q_2^2)^2
\right)
\right\}.
\label{eq:A0}
\eeq
The expressions for axial and tensor mesons are rather tedious, so we will not present them here.

\begin{figure}
\begin{center}
\begin{picture}(300,250)
\put(285,10){$-q_2^2,\ \mathrm{GeV}^2$}
\put(10,180){$\B^{(J)}(q_1^2,q_2^2)$}
\put(190,120){$J=0$}\put(250,40){$J=1$}\put(250,150){$J=2$}
\put(0,0){\epsfxsize=10cm \epsfbox{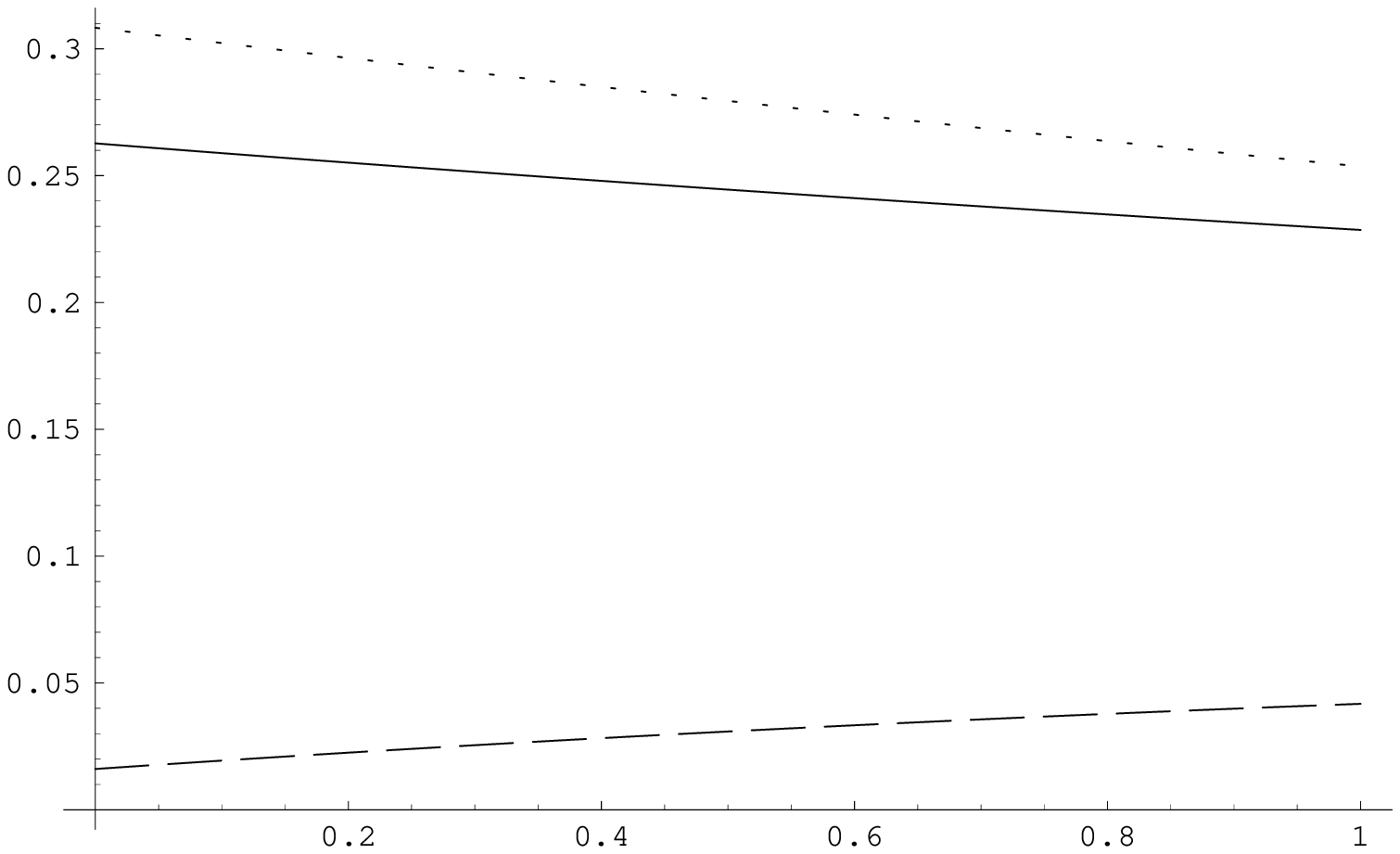}}
\end{picture}
\end{center}
\caption{$\B^{(J)}(q_1^2=-1\,\mathrm{GeV}^2, q_2^2)$\label{fig:BJ}}
\end{figure}

Squaring these amplitudes and averaging over the gluon polarizations we have
\beqa
\B^{(J)}(q_1^2,q_2^2) &=& \A^{(J)}_{\mu\nu}\A^{(J)*}_{\alpha\beta}g^{\mu\alpha}g^{\nu\beta},
\eeqa
where the explicit form of the functions $\B^{(J)}(q_1^2,q_2^2)$ is
\beq
\B^{(0)}(q_1^2,q_2^2) &=&
    \frac{8\pi\alpha_s^2|R'_\chi(0)|^2}{3M^3(q_1q_2)^4} \Big\{
9 M^8-24 q_1^2 M^6-24 q_2^2 M^6+22 q_1^4 M^4+22 q_2^4 M^4+
\nonumber \\ &+&
28 q_1^2 q_2^2 M^4 -
8 q_1^6 M^2-8 q_2^6
   M^2+8 q_1^2 q_2^4 M^2+8 q_1^4 q_2^2 M^2+q_1^8+q_2^8-
\nonumber \\ &-&
   4 q_1^2 q_2^6+6 q_1^4
   q_2^4-4 q_1^6 q_2^2
\Big\},
\label{eq:B0}\\
\B^{(1)}(q_1^2,q_2^2) &=&
    \frac{16\pi\alpha_s^2|R'_\chi(0)|^2}{M^3(q_1q_2)^4} \Big\{
    q_1^2 M^6+q_2^2 M^6-q_1^4 M^4-q_2^4 M^4-14 q_1^2 q_2^2 M^4-
\nonumber \\ &-&
    q_1^6 M^2-q_2^6 M^2+
    17 q_1^2 q_2^4 M^2+17 q_1^4 q_2^2 M^2+q_1^8+q_2^8-2 q_1^4 q_2^4
\Big\}
\label{eq:B1} \\
\B^{(2)}(q_1^2,q_2^2) &=& \frac{16\pi\alpha_s^2|R'_\chi(0)|^2}{3M^3(q_1q_2)^4} \Big\{
6 M^8-9 q_1^2 M^6-9 q_2^2 M^6+q_1^4 M^4+q_2^4 M^4+
\nonumber \\ &^+&
34 q_1^2 q_2^2 M^4+q_1^6 M^2+q_2^6 M^2-q_1^2 q_2^4 M^2-q_1^4 q_2^2 M^2+q_1^8+q_2^8-
\nonumber \\ &^-&
4 q_1^2 q_2^6+6 q_1^4 \
q_2^4-4 q_1^6 q_2^2
\Big\}.
\label{eq:B2}
\eeq
It can be seen that in the case $q_1^2=q_2^2=0$ we have $\B^{(1)}=0$, i.e. an axial $\chi$-meson can not be produced from the fusion of two massless gluons, exactly as the Landau-Yang theorem states. This is not true for scalar and tensor mesons, so the decays of these mesons into two gluons is possible. The widths of these decays can be expressed through the functions $\B^{(J)}$ by the relation
\beqa
\Gamma(\chi_{cJ}\to gg) &=&\frac{1}{2J+1}\frac{1}{2M}\frac{1}{2}\frac{1}{8\pi}\B^{(0,2)}(0,0),
\eeqa
that directly leads to eq. \eqref{eq:G0} and \eqref{eq:G2}. One other interesting property of $\B^{(J)}(q_1^2,q_2^2)$ functions is that in the case of axial meson we observe a strong dependence on the virtualities of the gluons, while for scalar and tensor mesons this dependence is not so crucial (see fig. \ref{fig:BJ}).

\section{Partonic subprocesses}

In this section we will consider the partonic subprocesses that give the contributions to production of charmonium $\Q$ in the hadronic experiments. At the leading order (LO) of the perturbation theory (i.e. $\sim\alpha_s^2$) the only possible reactions are the processes $gg\to\Q$ with $\Q=\chi_{c0}$ or $\chi_{c2}$. Vector charmonium production is forbidden by the charge parity conservation, and $\Q=\chi_{c1}$ cannot be produced because of the Landau-Yang theorem. However, from the experimental data we know, that the cross sections of $\chi_{c1}$ and $\chi_{c2}$ production are comparable  \cite{Alexopoulos:1999wp}. The only way to solve this problem is to consider next to leading order processes (NLO) in the strong coupling constant. For this reason we will deal with the following partonic processes:
\beq
a(k_1) b(k_2) &\to& \Q(p) c(k_3),
\label{eq:ggQg}
\eeq
where in the parenthesis the momenta of the particles are shown. In the case of proton-proton interaction we will consider the reactions
\beqa
gg\to\Q g, &\quad& qg\to\Q q,
\eeqa
where $q$ is the valence $u$ or $d$-quark. In the case of $p\bar p$-annihilation the quark-antiquark annihilation is also possible
\beqa
q\bar q &\to& \Q g.
\eeqa

Consideration of NLO in strong interaction constant solves also the other mentioned in the introduction problem. In the widely used collinear approximation the integrated over the transverse momentum $k_T$ distribution functions are used. As a result, it turns out that initial partons and, hence, final charmonium, do not possess any transverse momentum.

In the rest of this section we will consider in detail the mentioned above partonic reactions and give the analytical expressions for the corresponding cross sections.

\subsection{$g g \to \Q g$}

\subsubsection{$\Q=\JP$}

\begin{figure}
\begin{center}
\includegraphics{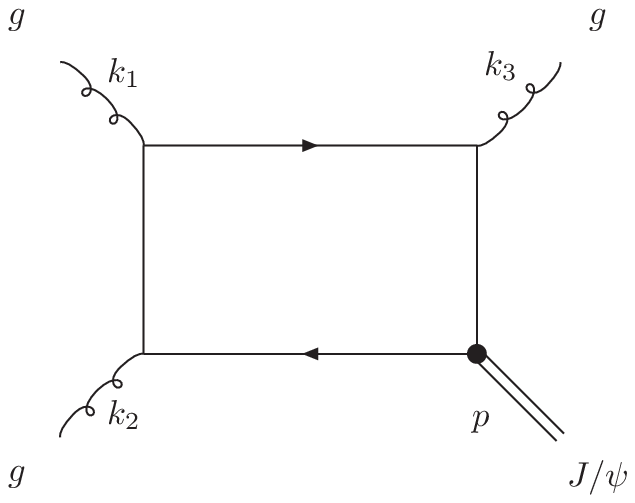}
\caption{$gg\to\JP g$ \label{diag:ggPsi}}
\end{center}
\end{figure}

Because of the charge parity conservation color-singlet $\JP$-state cannot be produced in the reaction $gg\to\JP$, so at LO it can be produced only with a gluon in the final state. One of the diagrams describing this process is shown in figure \ref{diag:ggPsi}, others can be obtain from it by gluon permutations. Using the formalism described in the previous section we have obtained the following expression for the partonic cross section:
\beqa
\frac{d\hsig(gg\to\JP g)}{d\hT} &=&
  \frac{10\pi\alpha_s^3}{9} \frac{M|R_\psi(0)|^2}{\hS^2}
  \frac{ (\hT^2+\hT\hU+\hU^2)\hS^2 + \hT\hU(\hT+\hU)\hS+\hT^2\hU^2}{(\hS+\hT)(\hS+\hU)(\hT+\hU)},
\eeqa
where the definitions for Mandelstam variables $\hS$, $\hT$ and $\hU$ are
\beqa
\hS&=&(k_1+k_2)^2,\quad \hT=(k_1-k_3)^2,\quad \hU=(k_1-k_2)^2.
\eeqa
These variables are not independent, the following relation between them exists:
\beqa
\hS+\hT+\hU &=& M^2.
\eeqa

Up to notations our result coincides with the result of work \cite{Gastmans:1987be}.

\subsubsection{$\Q=\chi_{c0,2}$}

\begin{figure}
\includegraphics{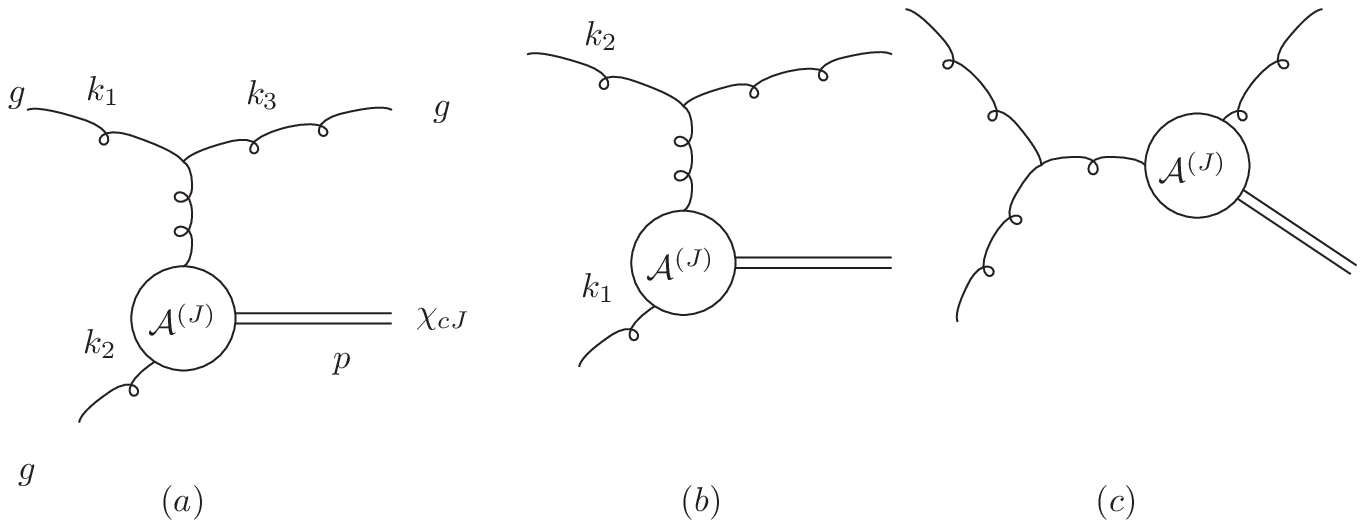}
\caption{$gg\to\chi_{cJ}g$\label{diag:ggChi}}
\end{figure}

In the case of $\chi_{c0,2}$-meson productions diagrams shown in figure  \ref{diag:ggPsi} give contribution to partonic cross section (obviously, $\JP$ in these diagrams should be replaced by $\chi_c$). It is also necessary to include diagram shown in figure \ref{diag:ggChi}.  The partonic cross sections of these processes are
\beq
\frac{d\hsig(gg\to\chi_{c0} g)}{d\hT} &=&
\frac{1}{\hT\hU}
\frac{4 \alpha _s^3 \pi  \left|R_\chi'(0)\right|^2}{\hat{s}^3 (\hat{s}+\hat{t})^4 (\
\hat{s}+\hat{u})^4 (\hat{t}+\hat{u})^4 M^3}
\Big[
9 (\hT+\hU)^4 M^{20}-
\nonumber \\ &-&
6 (\hT+\hU)^3 (9 \hT^2+14 \hU \hT+9 \hU^2) \
M^{18}+
\nonumber \\ &+&
(\hT+\hU)^2
   (153 \hT^4+492 \hU \hT^3+695 \hU^2 \hT^2+492 \hU^3 \hT+153 \hU^4) M^{16}-
\nonumber \\ &-&
   2    (\hT+\hU)^3 (135 \hT^4+393 \hU \hT^3+545 \hU^2 \hT^2+393 \hU^3 \hT+135 \
\hU^4)
   M^{14}+
\nonumber \\ &+&
   2 (162 \hT^8+1065 \hU \hT^7+3208 \hU^2 \hT^6+5852 \hU^3 \hT^5+7096 \
\hU^4 \hT^4+
\nonumber \\ &+&
5852 \hU^5 \hT^3+3208 \hU^6 \hT^2+1065 \hU^7 \hT+162 \hU^8) M^{12}-
\nonumber \\ &-&
   2 (135 \
\hT^9+966 \hU
   \hT^8+3215 \hU^2 \hT^7+6627 \hU^3 \hT^6+9351 \hU^4 \hT^5+
\nonumber \\ &+&
   9351 \hU^5 \
\hT^4+6627 \hU^6
   \hT^3+3215 \hU^7 \hT^2+966 \hU^8 \hT+135 \hU^9) M^{10}+
\nonumber \\ &+&
   (153 \hT^{10}+1170 \
\hU
   \hT^9+4249 \hU^2 \hT^8+9722 \hU^3 \hT^7+15548 \hU^4 \hT^6+
\nonumber \\ &+&
   18124 \hU^5 \hT^5+15548 \hU^6 \hT^4+9722 \hU^7 \hT^3+4249 \hU^8 \hT^2+1170 \hU^9 \hT
\nonumber \\ &+&
   153 \hU^{10}) M^8-  2 (27
   \hT^{11}+222 \hU \hT^{10}+885 \hU^2 \hT^9+2237 \hU^3 \hT^8+
\nonumber \\ &+&
   4001 \hU^4 \hT^7+5308 \hU^5 \hT^6+
5308 \hU^6 \hT^5+4001 \hU^7 \hT^4+2237 \hU^8 \hT^3+
\nonumber \\ &+&
885 \hU^9 \hT^2+222 \hU^{10} \hT+27 \hU^{11}) M^6+
   (9 \hT^{12}+90 \hU \hT^{11}+416 \hU^2 \hT^{10}
\nonumber \\ &+&
   1190 \hU^3 \hT^9+2394 \hU^4 \hT^8+3582 \hU^5 \hT^7+4090 \hU^6 \hT^6+
\nonumber \\ &+&
   3582 \hU^7 \hT^5+
   2394 \hU^8
   \hT^4+1190 \hU^9 \hT^3+416 \hU^{10} \hT^2+90 \hU^{11} \hT+
\nonumber \\ &+&
   9 \hU^{12}) M^4- 2 \hT \hU
   (\hT^2+\hU \hT+\hU^2)^2 (3 \hT^7+15 \hU \hT^6+37 \hU^2 \hT^5+
\nonumber \\ &+&
   55 \hU^3 \hT^4+   55 \hU^4 \hT^3+ 37 \hU^5 \hT^2+15 \hU^6 \hT+3 \hU^7) M^2+
\nonumber \\ &+&
   \hT^2 \hU^2 (\hT+\hU)^2 (\hT^2+\hU \hT+\hU^2)^4\Big],
\label{eq:ggChi0}
\eeq
\beq
\frac{d\hsig(gg\to\chi_{c2} g)}{d\hT} &=&
\frac{1}{\hT\hU}\frac{4 \alpha _s^3 \pi  \left|R_\chi'(0)\right|^2}{\hat{s}^3 (\hat{s}+\hat{t})^4 (\
\hat{s}+\hat{u})^4 (\hat{t}+\hat{u})^4 M^3}
\Big[
12 (\hT+\hU)^4 M^{20}-
\nonumber \\ &-&
24 (\hT+\hU)^3 (3 \hT^2+5 \hU \hT+3 \hU^2) M^{18}+
\nonumber \\ &+&
(\hT+\
\hU)^2
   (204 \hT^4+651 \hU \hT^3+880 \hU^2 \hT^2+651 \hU^3 \hT+204 \hU^4)M^{16}+
\nonumber \\ &+&
(-360
   \hT^7-1995 \hU \hT^6-4949 \hU^2 \hT^5-7428 \hU^3 \hT^4-7428 \hU^4 \hT^3-
\nonumber \\ &-&
4949 \hU^5
   \hT^2-1995 \hU^6 \hT-360 \hU^7) M^{14}+
\nonumber \\ &+&
   (432 \hT^8+2526 \hU \hT^7+6652 \hU^2 \hT^6+10877
   \hU^3 \hT^5+12640 \hU^4 \hT^4+10877 \hU^5 \hT^3+
\nonumber \\ &+&
   6652 \hU^6 \hT^2+
   2526 \hU^7 \hT+432   \hU^8) M^{12}+
\nonumber \\ &+&
(-360 \hT^9-2274 \hU \hT^8-6290 \hU^2 \hT^7-10647 \hU^3 \
\hT^6-13185 \hU^4 \hT^5-13185 \hU^5 \hT^4-
\nonumber \\ &-&
10647 \hU^6 \hT^3-6290 \hU^7 \hT^2-2274 \hU^8 \hT-360 \hU^9) M^{10}+
\nonumber \\ &+&
   (204 \hT^{10}+1455 \hU \hT^9+4328 \hU^2 \hT^8+7504 \hU^3 \hT^7+9232 \
\hU^4 \hT^6+9614 \hU^5 \hT^5+
\nonumber \\ &+&
9232 \hU^6 \hT^4+7504 \hU^7 \hT^3+4328 \hU^8 \hT^2+1455 \hU^9 \
\hT+204
   \hU^{10}) M^8+
\nonumber \\ &+&
   (-72 \hT^{11}-615 \hU \hT^{10}-2085 \hU^2 \hT^9-3878 \hU^3 \
\hT^8-4748 \hU^4 \hT^7-4678 \hU^5 \hT^6-
\nonumber \\ &-&
   4678 \hU^6 \hT^5-4748 \hU^7 \hT^4-3878 \hU^8 \
\hT^3-2085 \hU^9
   \hT^2-615 \hU^{10} \hT-72 \hU^{11}) M^6+
\nonumber \\ &+&
   (12 \hT^{12}+144 \hU \hT^{11}+616 \
\hU^2
   \hT^{10}+1345 \hU^3 \hT^9+1824 \hU^4 \hT^8+1806 \hU^5 \hT^7+
\nonumber \\ &+&
   1688 \hU^6 \hT^6+ 1806 \hU^7
   \hT^5+1824 \hU^8 \hT^4+1345 \hU^9 \hT^3+616 \hU^{10} \hT^2+144 \hU^{11} \
\hT+
\nonumber \\ &+&
12 \hU^{12}) M^4-
   \hT \hU (\hT^2+\hU \hT+\hU^2)^2 (12 \hT^7+60 \hU \hT^6+
\nonumber \\ &+&
   91 \hU^2
   \hT^5+49 \hU^3 \hT^4+49 \hU^4 \hT^3+91 \hU^5 \hT^2+60 \hU^6 \hT+
\nonumber \\ &+&
   12 \hU^7) M^2+2
   \hT^2 \hU^2 (\hT+\hU)^2 (\hT^2+\hU \hT+\hU^2)^4
\Big].
\label{eq:ggChi2}
\eeq
It should be noted that these cross sections diverge in the regions $\hT\to0$ and $\hT\to M^2-\hS$ (that is $\hU\to 0$). Such divergency is caused by the fact that in these regions the propagators of the virtual gluons shown in fig. \ref{diag:ggChi} become zero. As a result the cross sections of $\chi_{c0,2}$-production at $\hT\to0$ can be written in the factorized form:
\beq
\frac{d\hsig(gg\to\chi_{cJ}g)}{d\hT} &\underset{\hT\to0}{\approx}& \Phi_{g\to gg^*}(\hT)\frac{1}{\hT}\B^{(J)}(\hT,0).
\label{eq:fact}
\eeq
Here the first term
\beq
\Phi_{g\to gg^*}(\hT) &=& \frac{384\pi\alpha_s}{M^4}\frac{\left(\hS^2-\hS M^2+M^4\right)}{\hS(\hS-M^2)}
\label{eq:Phi}
\eeq
does not depend on the spin of the final charmonium and describes the splitting of the initial gluon into $gg^*$ pair, while the second term describes the production of $\chi_{cJ}$ meson [functions $B^{(J)}$ were presented earlier, formulas \eqref{eq:B0} --- \eqref{eq:B2}].

It is easy to understand the physical nature of this divergency. In the regions $\hT\to0$ and $\hU\to0$ the transverse momentum of the final charmonium
\beq
p_T &=& \sqrt{\frac{\hT\hU}{\hS}}
\label{eq:pT}
\eeq
tends to zero and it is impossible to distinguish the reaction \eqref{eq:ggQg} from reaction of $\chi_{c}$ production in the gluon pair fusion, that is the process $gg\to\chi_{c0,2}$. To avoid this divergency we will impose the following restriction on $p_T$:
\beqa
p_T &>& \Delta
\eeqa
and use the inverse geometrical size of the charmonium as the cutoff parameter $\Delta$:
\beqa
\Delta &\sim& \frac{1}{R_{c\bar c}} \sim m_c v.
\eeqa

It is interesting to mention, that the ratio of this divergent cross sections tends in the  region of small $p_T$ to a finite value, that is equal to
\beqa
\lim\limits_{p_T\to0}\left\{
  \left.\frac{d\hsig(gg\to\chi_{c2}g)}{dp_T} \right/ \frac{d\hsig(gg\to\chi_{c0}g)}{dp_T}
\right\}
  = \frac{2J_{\chi_2}+1}{2J_{\chi_0}+1}\frac{\Gamma(\chi_{c2})}{\Gamma(\chi_{c0})}=\frac{4}{3}.
\eeqa
This relation follows immediately from the eq. \eqref{eq:fact}, since in the range of small $p_T$ we have the production of $\chi_c$-mesons from two almost massless gluons. That is why the partonic cross sections are, up to spin factor, proportional to the widths of the decays of these mesons into a massless gluon pair.

\subsubsection{$\Q=\chi_{c1}$}

If $\Q=\chi_{c1}$ all shown in figures \ref{diag:ggPsi}, \ref{diag:ggChi} diagrams give contributions to the partonic cross section, and the expression for this cross section is
\beq
\frac{d\hsig(gg\to\chi_{c1} g)}{d\hT} &=&
\frac{12 \alpha _s^3 \pi  \left|R_P'\right|^2}{\hat{s}^2 (\hat{s}+\hat{t})^4 \
(\hat{s}+\hat{u})^4 (\hat{t}+\hat{u})^4 M^3}
\Big[
(\hT+\hU)^2 (\hT^2+\hU^2) M^{14}-
\nonumber  \\ &-&
4 (2 \hT^5+5 \hU \hT^4+6 \hU^2 \hT^3+6 \hU^3
   \hT^2+5 \hU^4 \hT+2 \hU^5) M^{12}+
\nonumber  \\ &+&
   (26 \hT^6+80 \hU \hT^5+115 \hU^2 \
\hT^4+120 \hU^3
   \hT^3+115 \hU^4 \hT^2+80 \hU^5 \hT+26 \hU^6) M^{10}-
\nonumber  \\ &-&
   2 (22 \hT^7+82 \hU \hT^6+141
   \hU^2 \hT^5+164 \hU^3 \hT^4+164 \hU^4 \hT^3+141 \hU^5 \hT^2+82 \hU^6 \hT+
\nonumber  \\ &+&
   22 \hU^7) M^8+(41 \hT^8+184 \hU \hT^7+378 \hU^2 \hT^6+510 \hU^3 \hT^5+
\nonumber  \\ &+&
   546 \hU^4 \hT^4+510 \hU^5 \hT^3+378 \hU^6 \hT^2+184 \hU^7 \hT+
\nonumber  \\ &+&
   41 \hU^8) M^6-2 (10 \hT^9+55 \hU \hT^8+136 \hU^2 \hT^7+218 \hU^3 \hT^6+265 \hU^4 \hT^5+265 \hU^5 \hT^4+
\nonumber  \\ &+&
   218 \hU^6 \
\hT^3+136 \hU^7
   \hT^2+55 \hU^8 \hT+10 \hU^9) M^4+(\hT^2+\hU \hT+\hU^2)^2 (4 \hT^6+22 \hU
   \hT^5+
\nonumber  \\ &+&
   37 \hU^2 \hT^4+36 \hU^3 \hT^3+37 \hU^4 \hT^2+22 \hU^5 \hT+4 \hU^6) M^2-
\nonumber  \\ &+&
2 \hT   \hU (\hT+\hU) (\hT^2+\hU \hT+\hU^2)^4
\Big].
\label{eq:ggChi1}
\eeq
It is clear that, contrary to $\chi_{c0,2}$ case, this cross section is finite in small $p_T$ region. This is explained by the Landau-Yang theorem, that forbids the production of $\chi_{c1}$-meson from two massless gluons. As a result, the squared matrix element of this reaction is proportional to the virtuality of the intermediate gluon (for example, $t$-channel gluon in fig.\ref{diag:ggChi}a), so this factor compensates the divergency, caused by the propagator.

\subsection{$g q \to \Q q$, $g \bar q \to \Q\bar q$}

\begin{figure}
\begin{center}
\includegraphics{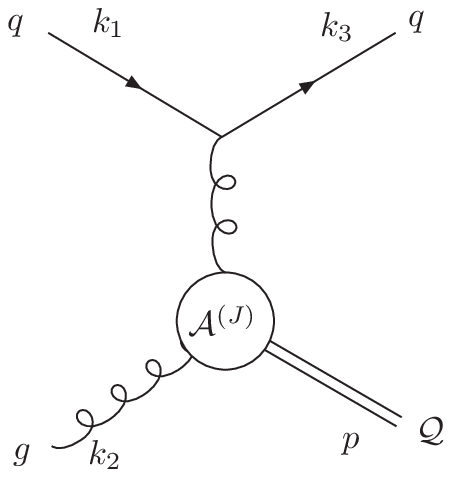}
\caption{$qg\to\Q q$ \label{diag:qg}}
\end{center}
\end{figure}

In this section we will consider charmonium production via scattering of the gluon on light (anti)quark. The corresponding diagrams are shown in figure \ref{diag:qg}. Only $\chi_c$ mesons can be produced in these reactions and the matrix element has the form
\beqa
\M(gq\to\chi_{cJ} q) &=&
    \frac{g_s}{q^2}\left(\lambda^a\right)_{ij}\delta_{ab}{\bar u}(k_3)\gamma^\mu u(k_1)\eps_2^\nu
    \A^{(J)}_{\mu\nu}(k_3-k_1,k_2),
\eeqa
where $i,j$ and $a,b$ are color indexes of quarks and gluons, $\eps_2^\nu$ is the polarization vector of the initial gluon, $\bar u(k_3)$ and $u(k_1)$ are the wave functions of quark and antiquark respectively, and $\A^{(J)}_{\mu\nu}$ are the vertices of the interaction of the gluon pair with the $\chi_c$-mesons, that were introduced earlier.

The partonic cross sections of the processes $qg\to\chi_{cJ}q$ have the form
\beqa
\frac{d\hsig(qg\to\chi_{c0} q)}{d\hT} &=&
\frac{1}{-\hT}\frac{8 \alpha _s^3 \pi  \left|R_P'\right|^2}{9 \hat{s}^2 M^3 \
\left(M^2-\hat{t}\right)^4}
\Big[
9 M^8-6 (3 \hat{s}+4 \hT) M^6+(18 \hat{s}^2+30 \hT \hat{s}+22 \hT^2) M^4-
\nonumber \\ &-&
2 \
\hT (6 \hat{s}^2+7 \hT
   \hat{s}+4 \hT^2) M^2+\hT^2 (2 \hat{s}^2+2 \hT \hat{s}+\hT^2)
\Big]
,\\
\frac{d\hsig(qg\to\chi_{c2} q)}{d\hT} &=&
\frac{1}{-\hT}\frac{16 \alpha _s^3 \pi  \left|R_P'\right|^2}{9 \hat{s}^2 M^3 \
\left(M^2-\hat{t}\right)^4}
\Big[
6 M^8-12 (\hat{s}+\hT) M^6+(12 \hat{s}^2+24 \hT \hat{s}+7 \hT^2) M^4-
\nonumber \\ &-&
2 \hT (6 \
\hat{s}^2+7 \hT
   \hat{s}+\hT^2) M^2+\hT^2 (2 \hat{s}^2+2 \hT \hat{s}+\hT^2)
\Big]
,\\
\frac{d\hsig(qg\to\chi_{c1} q)}{d\hT} &=&
-\frac{16 \alpha _s^3 \pi  \left|R_P'\right|^2}{3 \hat{s}^2 M^3 \
\left(M^2-\hat{t}\right)^4}
\Big[
(4 \hat{s}+\hT) M^4-2 (2 \hat{s}^2+3 \hT \hat{s}+\hT^2) M^2+
\nonumber  \\ &+&
\hT (2 \hat{s}^2+2 \hT \hat{s}+\hT^2)
\Big].
\eeqa
We see that, similar to gluon-gluon reactions, that cross section is finite for $\Q=\chi_{c1}$, and diverges  in the range of small $p_T$ for $\Q=\chi_{c0,2}$. Their ratio of the divergent cross sections is
\beqa
\lim\limits_{p_T\to0}\left\{
  \left.\frac{d\hsig(gq\to\chi_{c2}q)}{d\hT} \right/ \frac{d\hsig(gq\to\chi_{c0}q)}{d\hT}
\right\}
  =\frac{4}{3},
\eeqa
exactly equal to the same ratio in the gluon-gluon case. This fact is not surprising, since the nature of the divergencies of $gg\to\chi g$ and $gq\to\chi q$ is the same.

\subsection{$q\bar q\to \Q g$\label{sec:qq}}

If we consider charmonia production in $p\bar p$-annihilation, there are also valence antiquarks in the initial hadrons. In this case charmonia production via the quark-antiquark annihilation is possible. Experimental data says \cite{Corden:1997}, that the ratio of the $\JP$-production cross sections in $p\bar p$ and $pp$ reactions at the energy of 40 GeV is equal to
\beqa
\frac{\sigma(p\bar p\to\JP+X)}{\sigma(pp\to\JP+X)} &=& 6.
\eeqa
This is a clear sign of the significance of $q\bar q$ process at low energies.

This reaction is evidently cross-symmetric with respect to the process considered in the previous section, so the matrix elements of these processes can be expressed through each other:
\beqa
\left|\M(q\bar q\to\Q g)\right|^2 &=&  \left. \left|\M(gq\to\Q q)\right|^2\right|_{\hS\leftrightarrow\hT}.
\eeqa
With the help of this relation it is easy to obtain the following expressions for the partonic cross sections:
\beqa
\frac{d\hsig(q\bar q\to\chi_{c0} g)}{d\hT} &=&
\frac{64 \alpha _s^3 \pi  \left|R_P'\right|^2}{27 \hat{s}^3 M^3 \
\left(M^2-\hat{s}\right)^4}
\Big[
9 M^8-6 (4 \hat{s}+3 \hT) M^6+(22 \hat{s}^2+30 \hT \hat{s}+18 \hT^2) M^4-
\nonumber \\ &-&
2 \
\hat{s} (4 \hat{s}^2+7 \hT
   \hat{s}+6 \hT^2) M^2+\hat{s}^2 (\hat{s}^2+2 \hT \hat{s}+2 \hT^2)
\Big]
,\\
\frac{d\hsig(q\bar q\to\chi_{c1} g)}{d\hT} &=&
\frac{128 \alpha _s^3 \pi  \left|R_P'\right|^2}{9 \hat{s}^2 M^3 \
\left(M^2-\hat{s}\right)^4}
\Big[
(\hat{s}+4 \hT) M^4-2 (\hat{s}^2+3 \hT \hat{s}+2 \hT^2) M^2+
\nonumber  \\ &+&
  \hat{s} (\hat{s}^2+2 \hT \hat{s}+2  \hT^2)
\Big],
\\
\frac{d\hsig(q\bar q\to\chi_{c2} g)}{d\hT} &=&
\frac{128 \alpha _s^3 \pi  \left|R_P'\right|^2}{27 \hat{s}^3 M^3 \
\left(M^2-\hat{s}\right)^4}
\Big[
6 M^8-12 (\hat{s}+\hT) M^6+(7 \hat{s}^2+24 \hT \hat{s}+12 \hT^2) M^4-
\nonumber \\ &-&
2 \
\hat{s} (\hat{s}^2+7 \hT \hat{s}+6
   \hT^2) M^2+\hat{s}^2 (\hat{s}^2+2 \hT \hat{s}+2 \hT^2)
\Big].
\eeqa
These cross sections are finite in whole kinematically available region.

\section{Hadronic cross sections}

Let us now proceed to experimentally observable values and consider the process
\beqa
A(p_1) B(p_2) &\to& \Q(p) + X,
\eeqa
where $A$ and $B$ are the initial hadrons, $\Q=\JP$ or $\chi_{cJ}$, and in the parenthesis we show the particle momenta. The cross section of this reaction is expressed through the cross sections of considered above partonic reactions:
\beq
d\sigma^{AB}(\Q) &=& d\sigma(AB\to\Q+X) =\sum\limits_{a,b,c} d\sigma^{AB}_{ab}(\Q)=
  \sum\limits_{a,b}d\sigma(AB\to ab\to\Q+X)=
\nonumber \\ &=&
  \int dx_1 dx_2 f_{a/A}(x_1) f_{b/B}(x_2) d\hsig(ab\to\Q c).
\label{eq:hadr}
\eeq
Here we sum over partons $a$ and $b$, $x_{1,2}$ are the momentum fractions held by these partons, and $f_{a/A}(x_1)$, $f_{b/B}(x_2)$ are the distribution functions of the partons in the initial hadrons.

It is convenient to use other integration variables:
\beqa
x &=& x_1-x_2
\eeqa
and the squared energy of the partonic pair
\beqa
\hS &=& (k_1+k_2)^2 \approx x_1 x_2 s,
\eeqa
where $s=(p_1+p_2)^2$ and in the last equation we have neglected all masses except the mass of the final charmonium. The hadronic cross section is now written as
\beq
\sigma^{AB}_{ab}(\Q, p_T>\Delta) &=&
  \int\limits_{\Delta}^{ (s-M^2)/(2\sqrt{s})} dp_T
  \int\limits_{(p_T+\sqrt{p_T^2+M^2})^2}^s \frac{d\hS}{s} \frac{d\hsig(ab\to\Q c)}{dp_T}
\times\nonumber\\ &\times &
  \int\limits_{-(1-\hS/s)}^{1-\hS/s} \frac{dx}{\tilde x} f_{a/A}(x_1) f_{b/B}(x_2),
\label{eq:hadr2}
\eeq
where
\beqa
\tilde x &=& x_1+x_2 =\sqrt{x^2+\frac{4\hS}{s}},
\eeqa
the charmonium transverse momentum $p_T$ is defined in eq. \eqref{eq:pT}, and
\beqa
\frac{d\hsig}{dp_T} &=& \frac{p_T\sqrt{\hS}}{\sqrt{\left(\frac{\hS-M^2}{2\sqrt{\hS}}\right)^2-p_T^2}}
 \left[ \frac{d\hsig}{d\hT} + (\hT\leftrightarrow \hU) \right].
\eeqa
For $p_T<\Delta$ it is impossible to distinguish the reactions of the charmonium production with the emission of an additional parton and without such an emission. For this reason in small $p_T$ region we will consider only LO reactions
\beqa
p p &\to& g g \to \chi_{c0,2}
\eeqa
and the cross sections of these reactions are equal to
\beqa
\sigma^{AB}_{LO}(\chi_{cJ})=\frac{\pi^2}{64}\frac{\Gamma(\chi_{cJ}\to gg)}{M^3}\int\frac{dx}{\tilde x}f_{g/A}(x_1) f_{g/B}(x_2).
\eeqa

In our numerical estimates we used the distribution functions presented in the works \cite{Alekhin:2002fv} and \cite{Martin:2006qz}. Qualitatively these distributions give the same results, though for small energies  some quantitative difference is observed. In our paper only the results obtained with the help of \cite{Alekhin:2002fv} are presented. These distributions were calculated at the scale
\beqa
Q^2=3\,\mathrm{GeV}^2
\eeqa
and other parameters are equal to
\beqa
\alpha_s=0.3, \quad M=3\,\mathrm{GeV}, \quad\Delta^2=0.3\,\mathrm{GeV}^2\\
|R_S(0)|^2=0.81\,\mathrm{GeV}^3,\quad |R_P'(0)|^2=0.075\,\mathrm{GeV}^5.
\eeqa

\begin{figure}
\begin{picture}(500,250)
\put(400,75){$\sqrt{s}$, GeV}
\put(45,240){$\sigma$, $\mu b$}
\put(0,0){\epsfxsize=15cm \epsfbox{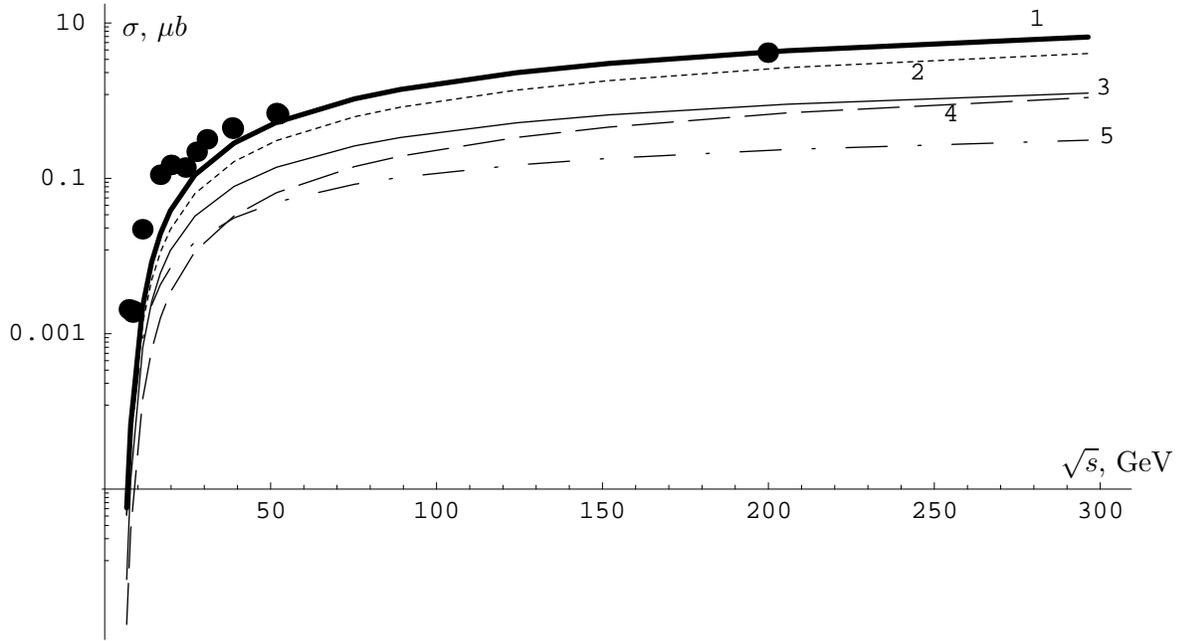}}
\end{picture}
\caption{
Cross sections of the charmonium production in proton-proton reaction for different values of c.m. energy. 1 --- total $\JP$ production, 2 --- $\JP$ production through the radiative $\chi_{c2}$ decay, 3 --- direct $\JP$, 4 --- $\JP$ produced in radiative $\chi_{c1}$ decay, 5 --- collinear contribution to $\chi_{c2}$ production.
\label{fig:Sigma}}
\end{figure}

In figure \ref{fig:Sigma} we show the dependence of $\JP$ production through different processes on total energy. The bold line (label 1) shows  the summed over all processes
\beqa
\sigma &=& \sigma^{pp}_{gg}(\JP)+\Br(\chi_{c1}\to\JP\gamma)\sigma^{pp}(\chi_{c1})+\Br(\chi_{c2}\to\JP\gamma)\sigma^{pp}(\chi_{c2}),
\eeqa
dotted line (label 2) --- $\JP$ production from radiative $\chi_{c2}$ decay, direct $\JP$ production (thin solid line, label 3); $\JP$-production from radiative $\chi_{c1}$ decay (dashed line, label 3) and collinear $\chi_{c2}$ contribution (dash-dotted line, label 5). Dots are the experimental results, presented in \cite{Adler:2003qs}. It is seen that our calculations give satisfactory agreement with experimental data in wide interval of energies, though for small energies there is some difference. This difference is caused by the fact, that our cross sections of direct $\JP$ and $\chi_{c1}$ productions are smaller, than the experimental results. Another possible reason of the disagreement is that for small energies it is not valid to neglect the proton mass in comparison with the energy of the reaction. This mass leads to the decrease of the partonic energy and increases the cross section and can be taken into account shifting the theoretical curves along the horizontal direction. At high energy this effect is not so crucial, as it is clearly seen from figure \ref{fig:Sigma}.

\begin{figure}
\begin{center}
\begin{picture}(300,200)
\put(285,10){$\sqrt{s}$, GeV}
\put(10,180){$R_{qg/gg}$}
\put(0,0){\epsfxsize=10cm \epsfbox{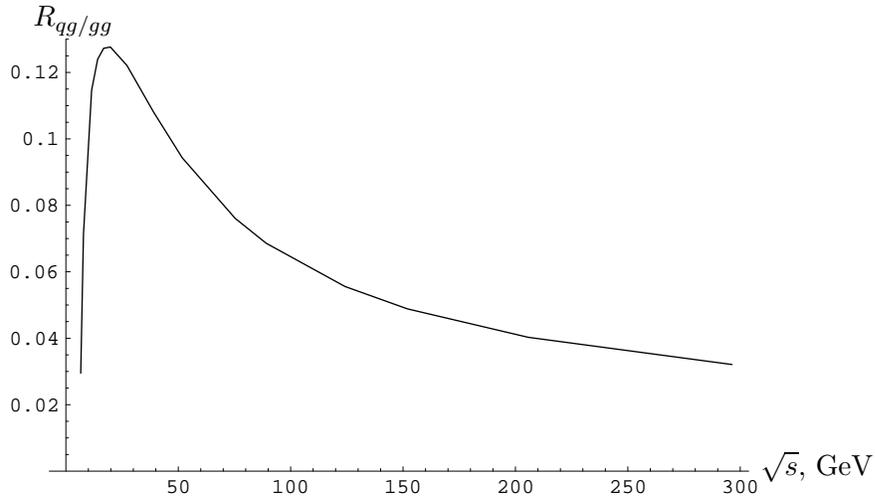}}
\end{picture}
\end{center}
\caption{The role of the quark-gluon subprocesses versus the energy of the reaction\label{fig:QGRatio}}
\end{figure}

We would like to stress, that the relative contributions of the mechanisms of $\JP$ production depend strongly on the energy of the reaction. For example, in figure \ref{fig:QGRatio} we show the dependence of the ratio
\beqa
R_{qg/gg} &=& 2\frac{
  \Br(\chi_{c1}\to\JP\gamma)[\sigma_{gu}^{pp}(\chi_{c1})+\sigma_{gd}^{pp}(\chi_{c1})]+
  \Br(\chi_{c2}\to\JP\gamma)[\sigma_{gu}^{pp}(\chi_{c2})+\sigma_{gd}^{pp}(\chi_{c2})]}
{
  \sigma_{gg}^{pp}(\JP)+\Br(\chi_{c1}\to\JP\gamma)\sigma_{gg}^{pp}(\chi_{c1})+\Br(\chi_{c2}\to\JP\gamma)\sigma_{gg}^{pp}(\chi_{c2})
},
\eeqa
on the energy. This ratio describes the role of quark-gluon subprocesses in the production of $\JP$-meson. In is seen, that it depends strongly from the energy, but its value is suppressed in the whole shown range.

Our results for $\chi_{c1}$ and direct $\JP$ production cross sections are somewhat smaller, than the experimental values. From experimental data \cite{Alexopoulos:1999wp} we know, that the ratio of $\chi_{c1}$ and $\chi_{c2}$ production cross sections is equal to
\beqa
\frac{\sigma(\chi_{c1})}{\sigma(\chi_{c2})} &\approx& 0.3,
\eeqa
while our calculations give the value approximately 3 times smaller. We think that the source of this difference is that only one of the producing $\chi_{c1}$ gluons was virtual. According to figure \ref{fig:BJ} the axial charmonium is very sensitive to the virtualities of the gluons, so this effect will increase the cross section of its production. The $\chi_{c0}$ and $\chi_{c2}$ production cross section would not change significantly. As for the direct $\JP$, in \cite{Clavelli:1984ee} it was shown, that the process $qg\to\JP q g$ almost doubles the value of its production cross section. It should be noticed, however, that the collinear singularity, appearing in this process, was removed by introducing the mass of the quark, and the value of this mass was chosen to be $m_q=5$ MeV. We think, that this value is too small, and hence, the role of $qg\to\JP q g$ process is overestimated in \cite{Clavelli:1984ee}.

\begin{figure}
\begin{center}
\begin{picture}(300,200)
\put(290,10){$\sqrt{s}$, GeV}
\put(0,180){$\frac{\sigma(p\bar p\to\JP+X)}{\sigma(pp\to\JP+X)}$}
\put(0,0){\epsfxsize=10cm \epsfbox{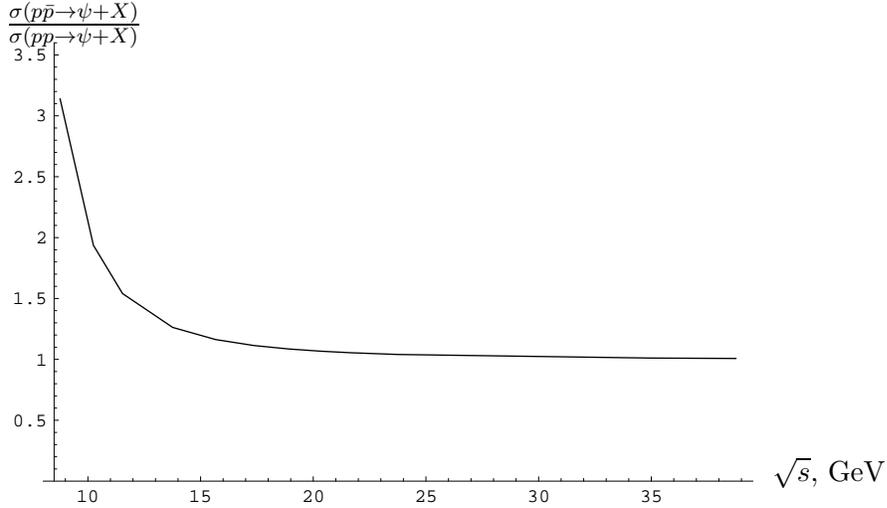}}
\end{picture}
\end{center}
\caption{
The ratio of the cross sections of $\JP$ production in $p\bar p$ and $pp$ reactions
\label{fig:PbPP}}
\end{figure}

In the case of $p\bar p$ annihilation there are also valence antiquarks in the initial hadrons. That is why the cross section of $p\bar p\to\Q+X$ reaction will receive the contributions from the quark-antiquark subprocesses, that were considered in section 3.3. In figure \ref{fig:PbPP} we show the dependence on the energy of the ratio of the cross sections $\sigma(p\bar p\to\JP X)$ and $\sigma(pp\to\JP X)$. It can be seen, that this ratio is significant for small energies and tends to unity for higher ones. The reason for this is that the quark-antiquark cross sections are suppressed by the propagator of the $s$-channel gluon. Such behavior of this ratio is observed in the experiment.

\begin{figure}
\begin{center}
\begin{picture}(300,200)
\put(290,10){$p_T$, GeV}
\put(0,180){$\frac{\Br(\JP\to e^+e^-)}{p_T}\frac{d\sigma}{dp_T},\,\frac{\mathrm{nb}}{\mathrm{GeV}}$}
\put(0,0){\epsfxsize=10cm \epsfbox{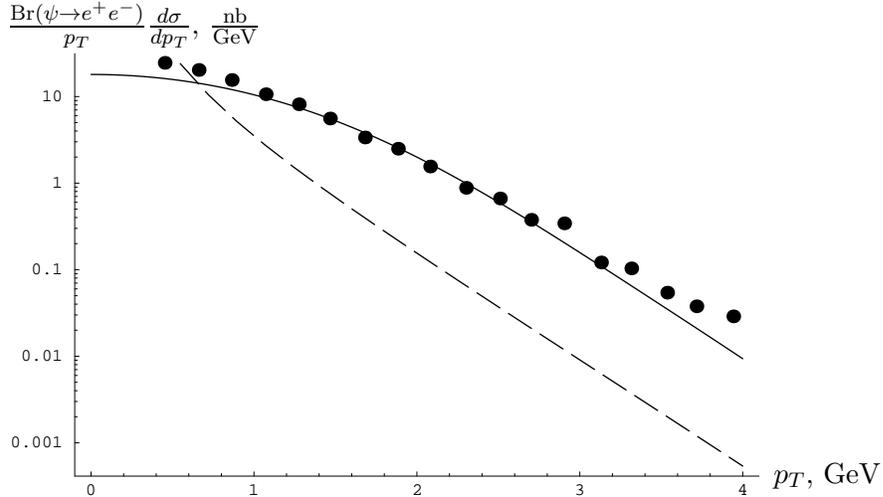}}
\end{picture}
\end{center}
\caption{
Transverse momentum distribution of the total $\JP$-production cross section at energy $\sqrt{s}\approx27$ GeV. Dashed line --- our cross section, solid line --- smeared distribution, dots --- experimental values \cite{Siskind:1979ax}.\label{fig:SigmaPT}}
\end{figure}

Staring from formula \eqref{eq:hadr} it is easy to obtain the distribution of the charmonia production cross section over the transverse momentum. In figure \ref{fig:SigmaPT} with the dashed line we show this distribution for energy $\sqrt{s}=27$ GeV and the experimental data, taken from \cite{Siskind:1979ax}. It is clear, that our technique gives too fast decrease of the distribution with the rise of transverse momentum. In order to remove this disagreement one can take into account the transverse motion of the initial gluons in the proton. This method is used in the so called $k_T$ factorization to explain the high energy results. For low energies, unfortunately, it cannot be used, since the unintegrated distribution functions at these scales are poorly known. We will take into account the transverse motion of initial gluons by phenomenological correction \cite{Clavelli:1984ee}
\beq
\frac{d\sigma}{dp_T} &\to& \frac{A}{4\pi\sigma^2}\int d^2q_T \frac{d\sigma}{dq_T} e^{-(\mathbf{p}_T-\mathbf{q}_T)^2/4\sigma^2},
\label{eq:smear}
\eeq
where
\beq
A &=& 3,\qquad \sigma=0.5\,\mathrm{GeV}.
\label{eq:smearPar}
\eeq
The corrected distribution is shown in figure \ref{fig:SigmaPT} with solid line.

\subsection{$\E=70$ GeV}

In this section we will consider in more details charmonium production at $\E=70$ GeV.

\begin{figure}
\begin{center}
\begin{picture}(300,200)
\put(290,10){$p_T$, GeV}
\put(0,180){$\frac{1}{p_T}\frac{d\sigma}{dp_T},\,\frac{nb}{GeV}$}
\put(0,0){\epsfxsize=10cm \epsfbox{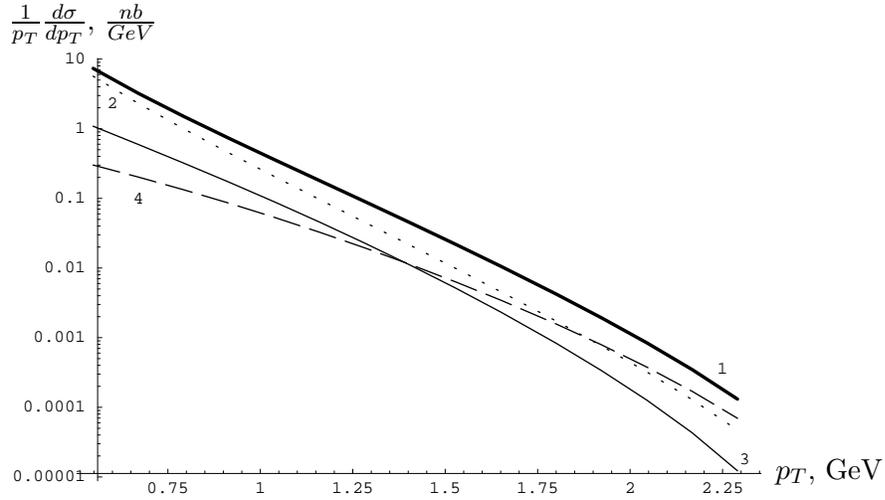}}
\end{picture}
\end{center}
\caption{
Transverse momentum distributions of $\JP$ production. Thick solid line --- total cross sections; thin solid line --- direct $\JP$ production; dashed line --- $\JP$ form radiative $\chi_{c1}$ decay; dotted line --- $\JP$ form radiative $\chi_{c2}$ decay \label{fig:pT}}
\end{figure}

In figure \ref{fig:pT} the $p_T$ distributions of $\JP$ productions form different mechanisms are shown: direct $\JP$ production (thin solid line, label 3), $\JP$ form radiative $\chi_c$ decays ($\chi_{c1}$ --- dashed line, label 4; $\chi_{c2}$ --- dotted line, label 2) and total cross section (thick solid line, label 1). We have not performed any smearing, since the value of the parameters $\sigma$ and $A$ is unknown.

\begin{figure}
\begin{center}
\begin{picture}(300,200)
\put(285,10){$x_F$}
\put(130,185){$\frac{d\sigma}{dx_F},\,ναν$}
\put(0,0){\epsfxsize=10cm \epsfbox{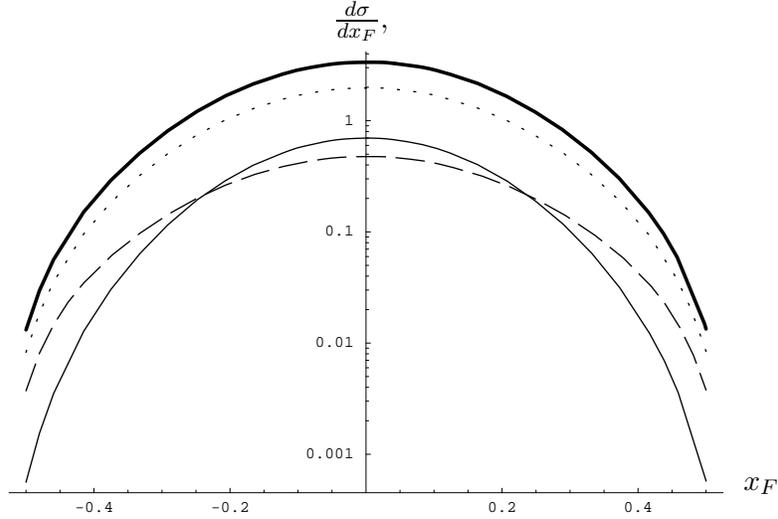}}
\end{picture}
\end{center}
\caption{$x_F$ distributions. Solid bold line --- total cross section; thin solid line --- direct $\JP$ production; dashed and dotted lines --- $\JP$ from $\chi_{c1}$ and $\chi_{c2}$ respectively
\label{fig:xF}}
\end{figure}

The other widely used kinematical variable is the longitudinal charmonium momentum $p_L$. It is, however, more convenient to work with dimensionless variable
\beqa
x_F &=& \frac{2p_L}{\sqrt{s}},
\eeqa
that can be expressed through the introduced earlier values:
\beq
x_F &=& \frac{\hS+M^2}{2\hS}x+\frac{\hS-M^2}{2\hS}\tilde{x} z,
\label{eq:xF}
\eeq
where
\beqa
z&=&\cos\hat\theta_{13}=1+\frac{2\hT}{\hS-M^2},
\eeqa
and $\hat\theta_{13}$ is the angle between the momenta ${\bf k}_1$ and ${\bf k}_3$, measured in partonic rest frame. In figure \ref{fig:xF} we show $x_F$-distributions of the cross sections for $\JP$ and $\chi$ states. It can be seen, that radiative $\chi_{c2}$ decay gives the main contribution to $\JP$ production cross section.

\begin{figure}
\begin{center}
\begin{picture}(300,200)
\put(285,10){$x_F$}
\put(135,185){$x_{1,2}$}
\put(275,160){$x_1$}
\put(275,50){$x_2$}
\put(0,0){\epsfxsize=10cm \epsfbox{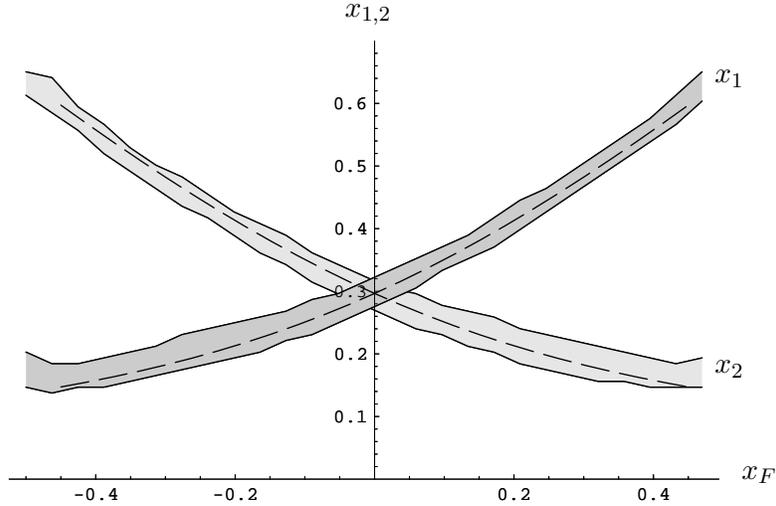}}
\end{picture}
\end{center}
\caption{
Correspondence between experimentally observed variable $x_f$ and momentum fractions $x_{1,2}$
\label{fig:x12}}
\end{figure}

For the determination of the gluon distribution functions from the experimental data one needs to know the momentum fractions of these partons. These fractions, unfortunately, cannot be determined directly. So, it is necessary to know what values of $x_{1,2}$ correspond to specific value of the variable $x_F$. If the energy of the partonic reactions coincides precisely with charmonium mass (that is we deal with the resonant charmonium production), than only the first term of relation \eqref{eq:xF} gives nonzero contribution and we have one to one correspondence between $x_{1,2}$ and $x_F$:
\beq
x_{1,2} &=& \frac{1}{2}\left(\sqrt{x_F^2+\frac{4\hS}{s}}\pm x_F\right).
\label{eq:x12}
\eeq
If the energy of the partonic reaction is larger than the charmonium mass, this correspondence is ambiguous. We know, however, that the states with large partonic energies are suppressed by the partonic luminocities (that is the probability to observe partons with large momentum fractions  $x_1$ and $x_2$), so the contribution of these states into cross sections will not be significant. For this reason the errors caused by the mentioned ambiguous will be small. In figure \ref{fig:x12} we show the correspondence between $x_{1,2}$ and $x_F$ and the errors in the determination of these variables. The dashed lines correspond to the relation \eqref{eq:x12} with $\hS=1.3M^2$. The $x_1$ error was determined from the "half-width" condition, that is
\beqa
\frac{d^2\sigma}{dx_Fdx_1} &>& \left.\frac{1}{2}\frac{d^2\sigma}{dx_Fdx_1}\right|_{\mathrm{max}}
\eeqa
(the same for $x_2$). We would like to mention, that these errors have nothing to do with the experimental uncertainties and cannot be reduced by the increase of the experimental precision.

\section{Spin asymmetries}

The ultimate goal of the PAS-CHARM experiment is the determination of the spin distribution functions of the partons in the polarized nuclon $f(\lambda,x)$. Here $\lambda=+1$ corresponds to the case of the parton spin aligned along the spin of the parent nucleon, while $\lambda=-1$ corresponds to opposite alignment of these spins. For the determination of these functions it is proposed to consider the spin asymmetry
\beq
A_{LL} &=& \frac{\sigma(\uparrow\uparrow)-\sigma(\uparrow\downarrow)}{\sigma(\uparrow\uparrow)+\sigma(\uparrow\downarrow)},
\label{eq:ALL}
\eeq
where symbols $\uparrow$ and $\downarrow$ in the $\JP$-production in the proton-proton reaction cross section correspond to the cases with proton polarized along the direction of its motion or against this direction.

\begin{figure}
\begin{center}
\begin{picture}(300,200)
\put(285,95){$p_T$, GeV}
\put(0,180){$\hat a_{LL}\left(\frac{\Delta G}{G}\right)^2$}
\put(0,0){\epsfxsize=10cm \epsfbox{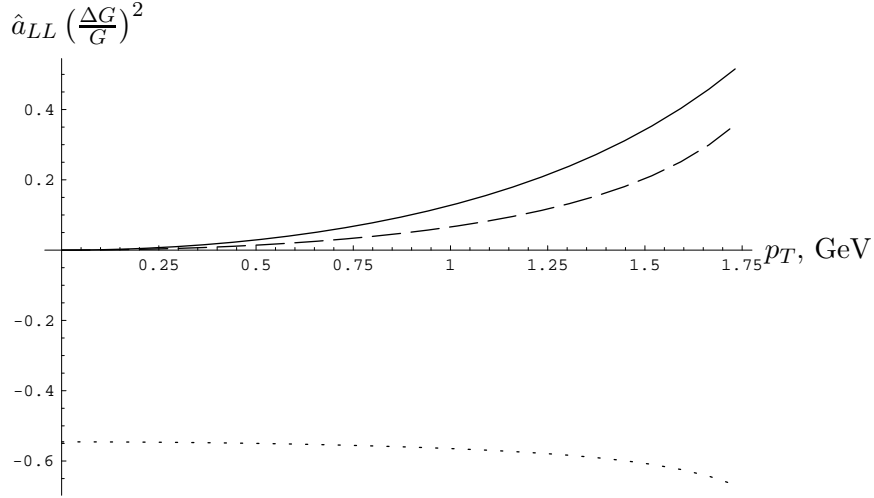}}
\end{picture}
\end{center}
\caption{
$p_T$ distributions of the partonic spin asymmetry $\hat a_{LL}$ for reactions $gg\to\JP g$ (solid line), $gg\to\chi_{c1}g$ (dashed line) and $gg\to\chi_{c2}g$ (dotted line)
\label{fig:aLL}}
\end{figure}

The explicit expression for asymmetry \eqref{eq:ALL} is \cite{Usubov:1996dz}
\beqa
A_{LL} &=& \hat a_{LL} \frac{\Delta f(x_1)}{f(x_1)}\frac{\Delta f(x_2)}{f(x_2)},
\eeqa
where $\Delta f(x)=f(+,x)-f(-,x)$ and the partonic spin asymmetry is connected with the cross sections of charmonium production from two partons with the Helicities $\lambda_1$ and $\lambda_2$:
\beqa
\hat a_{LL} &=& \frac{d\hsig(++)-d\hsig(+-)}{d\hsig(++)+d\hsig(+-)}.
\eeqa
Using the helicity cross sections of charmonium production in gluon-gluon annihilation \cite{Gastmans:1987be} and the experimental constraints on the ratio
\beqa
\frac{\Delta G}{G} &=& \frac{\Delta f_{g/p}(x)}{f_{g/p}(x)}
\eeqa
that reaches its maximum value 0.9 for $x\approx0.3$  \cite{Mallot:2006mc} it is easy to obtain the estimate for asymmetry $A_{LL}$ at energy 70 GeV. In figure \ref{fig:aLL} we show the $p_T$ distributions of the partonic asymmetries $\hat a_{LL}$. It is clearly seen, that different charmonia states give the opposite sign contributions into this asymmetry, so averaging over these states leads to significant decrease of the asymmetry. So it is necessary to determine the type of the experimentally observed charmonium meson and separate the contributions from directly produced $\JP$ and $\JP$ from radiative $\chi_{c1,2}$ decays.

\section{Conclusion}

In the proposed experiment SPASCHARM it is planned to measure the polarization distribution functions of quarks and gluons in the nucleon. For this the detailed analysis of different modes of charmonia production in the hadronic experiments is necessary. The collinear approximation, that is widely used in parton model, suffers from a number of serious drawbacks, that make it impossible to use this approximation in our case. First of all, this approach gives no information on the distributions over the transverse momentum of the final charmonium. These distributions give rich information on the charmonia production mechanisms. The second drawback is that at leading order in the strong coupling constant only $\chi_{c0,2}$-mesons can be produced, while the experimental data gives large direct $\JP$ and $\chi_{c1}$ signals.

The natural way to solve these difficulties is to consider the third-order processes in the strong coupling constant. This is the topic of our article. We have shown that the consideration of the higher order processes increases significantly the number of mechanisms of charmonia production. In addition to $\chi_{c0,2}$ production in gluon-gluon fusion, that is possible in the leading order, there appears also a production of direct $\JP$ and $\chi_{c1}$ and production of $P$-wave charmonia in quark-gluon interaction $gq\to\chi_{cJ} q$. In the case of proton-antiproton annihilation the production of $\chi_c$-mesons in the process $q\bar q\to\Q g$ is possible, that explains the experimentally observed excess of charmonia production in proton-antiproton annihilation at low energies in comparison with the proton-proton scattering.

The results, presented in our paper show, that this approach gives the cross section of total $\JP$ production in proton-proton interaction, that describes well the experimental data in a large range of energies, and qualitatively  describes the $\chi_{c1}$ production and distributions over the transverse momentum of the final charmonia. There is, however, no quantitative coincidence. Our values for $\chi_{c1}$ production cross section are about 3 times smaller, than the experimental values, and $p_T$ distributions fall with the transverse momentum faster, than it is observed experimentally. It is known from the theoretical and experimental analysis of TEVATRON data, that these contradictions can be removed with the help of $k_T$-factorization. Unfortunately, For low energies (for example the energies of U70) it is hard to use this approach, so we have introduced phenomenological corrections.

The presented in our paper detailed analysis of different mechanisms of charmonia production in proton-proton interaction at the energy of the initial proton equal to 70 GeV shows, that the relative contributions of these mechanisms depend strongly on the transverse momentum of the final charmonium and other kinematical variables.

In the last section of our article we have given the expression for spin asymmetries of partonic processes of charmonium production in gluon-gluon fusion. These asymmetries, with the analogous values for quark-gluon subprocesses, are necessary for the determination of the polarization distribution functions of quarks and gluons in proton. From our results it is clearly seen, that the contributions of different charmonia states into this asymmetry compensate each other. For this reason it is necessary do separate directly produced $\JP$ mesons and $\JP$-mesons produced via radiative $\chi_{c1,2}$-decays.

The authors would like to thank A.N. Vasiliev and V.V. Braguta for useful discussions. Our work was supported in part by Russian Foundation of Basic Research under grant 07-20-00417a.

\end{document}